\documentstyle[pra,aps,preprint]{revtex}

\begin{document}
\title{
Quantum Limit on Computational Time and Speed}
\author{A. K. Pati$^{1,2,3}$, S. R. Jain$^{4}$, A. Mitra$^{5}$, R. 
Ramanna$^{6}$}
\address{$^{1}$ Institute of Physics, Sainik School Post, Bhubaneswar-751005,
Orissa, India}
\address{$^{2}$ Theoretical Physics Division, BARC, Mumbai-400085, India}
\address{$^{3}$School of Informatics, University of Wales, 
Bangor LL 57 1UT, UK}
\address{$^{4}$ NPD, Vande-Graff Building, BARC, 
Mumbai-400085,  India}
\address{$^{5}$ NRL, BARC, Mumbai-400085,  India}
\address{$^{6}$National Institute of Advanced Studies, Indian
Institute of Science Campus, Bangalore 560012, India}
%\footnote{email:apati@apsara.barc.ernet.in}

\date{July 6, 2001}
\maketitle
\def\ra{\rangle}
\def\la{\langle}
\def\ver{\arrowvert}
\begin{abstract}
We investigate if physical laws can impose limit on computational time
and speed of a quantum computer built from elementary particles. We show 
that the product of the speed and the running time of a quantum computer 
is limited by the type of fundamental interactions present inside the system. 
This will help us to decide as to what type of interaction should be allowed 
in building quantum computers in achieving the desired speed.

\end{abstract}

\vskip .5cm

\vskip 1cm

%\newpage
%\begin{multicols}{2}

\vskip .5cm

Any computer, be it classical or quantal, stores information in 
the physical states and processes the same according to the laws of 
physics \cite{rpf}. Last century has witnessed tremendous growth of 
computational power in solving complex problems and in optimising 
computational resources. In the saga of information age it is
impetuous to know the ultimate limitations on the speed and the
maximal running time or the life-time of a computer. Since a computer 
is a physical device made up of basic elements, its underlying
dynamics, its ability, its performance and life 
time will depend on the laws of physics. The present day
computers store information in macroscopic elements and process
information according to the laws of
classical physics \cite{rpf}. For example, all the electronic
computers employ silicon memory chips and are
controlled by electrostatic or electromagnetic interactions. However, 
the quantum information revolution has offered
us possibility of realizing computers that can be built out of a
collection of {\em interacting} two-state
systems \cite{bd,nc,david}. Using the quantum features such as linear
superposition and quantum entanglement, it is hoped
that quantum computers can gain some extra power as compared to 
classical computers \cite{nc,david}. For example,
one can factorise an integer in polynomial time \cite{shor} and search
a data base in square root number of steps \cite{lkg} in a quantum
computer. In practical realisation of quantum computers a crucial
factor is that the interaction between
subsystems should be controlled in such a way that the speed and life
time of the computer should be increased.
Also the interaction between the computer and external world should be 
as small as possible to avoid decoherence \cite{zurek}.

It is often the case that quantum mechanical and relativity principles 
together with
the existence of fundamental constants in nature such as Planck's
constant $\hbar$, speed of light $c$ and universal Gravitational
constant $G$ some times provide important bound on the physical
quantities of interest. For example, it was shown that there is a
limit on the maximal acceleration of a quantum particle and accurate
clocks are those which can be accelerated maximally \cite{akp}. The
maximal acceleration was found to be the Planck acceleration $a_p=
\sqrt{\frac{\hbar G}{c^3}}$. Similarly, there is a limit on the power
radiated by charged particles in external force field \cite{akp1}.
Therefore, it is natural to seek if the ability of a computer is
limited by laws of physics, existence of fundamental constants and 
fundamental forces in the universe. Recently, there have been attempts 
to put limits on the speed and memory capacity of a computer.
Margolus and Levitin \cite{ml} have shown that the number of elementary
operations that a physical system can perform per second is limited by 
$\frac{2E}{\pi \hbar}$, where $E$ is average energy of the system.
Remarkably, Lloyd \cite{sl} has found that the speed of a computer is 
limited by the energy available to the machine and memory (the total
number of bits available for a physical system to process information) 
is by its entropy. Ng \cite{ng} has argued that the product of 
speed and the clock rate of a computer 
is limited by the reciprocal of squared Planck time $t_{p}^2= c^5/ \hbar G$. 
This bound arises again from laws of quantum physics and gravitation.
However, bounds on time and speed has not been shown to depend on the type
of fundamental interactions present in the universe.

In recent years one of the important goals is to enhance the speed and
the life time (or running time) of a computer. In view of the fact 
that physical world is ultimately quantum mechanical, there have 
been momentous developments in imagining quantum computers.
%which would promisingly solve some hard problems efficiently. 
To build a quantum computer one 
needs to know what type of interactions would be most appropriate
among its subsystems.
In this paper we ask the question: In quantum world, is there a limit
on the running time (or life time) and speed of a computer arising 
due to the fundamental {\em physical interactions} between the subsystems? 
In nature, we know that there are foure types of fundamental
interactions, namely, (i) strong interaction,
(ii) weak interaction, (iii) electromagnetic interaction and 
(iv) gravitational interaction. Also it is known that fundamental
interactions between elementary particles carrying information can be
used for quantum logic operations \cite{cz,sl1,dbe}. 
If a quantum computer is built from 
elementary particles subject to these interactions, then its
speed and running time should depend on the type of interaction. 
By combining principles of quantum theory, relativity and particle 
physics empirical systematics, we show that the product of running 
time $T$ and  the 
speed $v$ is upper bounded by $2^n/n$, where $n$ is an integer 
depending on the type of fundamental interaction 
present in the quantum system. This can be interpreted as a 
complimentary relation between speed and time of a quantum computer 
for a given natural interaction. 

First, let us note that quantum computers can be built out of
collection of logical two-state systems (called qubits) with suitable 
physical interactions (from above four fundamental forces) allowed by
laws of nature. In principle, we can utilise elementary particles with 
two distinct quantum states as our qubits. For
example, in case of strong interaction, we can identify neutron and 
proton as two distinct states of a single neucleon with iso-spin
interaction. Hence, in the presence of iso-spin interaction a neucleon can be
regarded as a qubit. In case of weak interaction, one has to look for
elementary particles that are subjected to weak forces. In this
case there is a well-known phenomena of neutrino
oscillation. By coherence properties \cite{knw} one can be regarded 
it as a qubit. In case of
electromagnetic interaction, there are numerous
examples. Simplest example, here, is that in the presence of
electromagnetic laser pulse, a two-state atom
under goes oscillation between its ground and excited states and this
can be regarded as a qubit \cite{nc}. Or we could think of spin of 
nuclei being manipulated using magnetic field such as in NMR devices. In
case of gravitational interaction, in principle, one can think of a 
quantum system in a superposition of two distinct mass states and if it is 
allowed that would constitute a qubit. Once one 
identifies qubits, then using the
desired one-qubit and two-qubits interaction \cite{dpd} one can construct 
universal logic gates to manipulate the collection of quantum systems.

Suppose, we have a quantum computer made out of elementary two-state
systems with any one of the four fundamental interaction between
them. Let the composite quantum system consists of ${\cal N}$
elementary subsystems each having mass $m$. Let $M$ be the rest mass 
and $\Psi(X)$ be the initial state of the quantum computer. Let us 
assume that the wavefunction has a linear spread $\delta X$ at an
initial time. By the Heisenberg's uncertainty relation its initial
momentum spread will be of the order of $\hbar/\delta X$.
Therefore, if the computer is allowed to evolve for a time $t$, the 
position of the wavepacket will be given by
\begin{equation}
\delta X(t) \simeq \delta X + \frac{\hbar t}{M \delta X} + 
(\delta X)_{\mbox{\small int}},
\end{equation}
where the last term gives the spread due to interaction with the surroundings.
We are considering the spread along one dimension but it applies
equally well to other dimensions.
We wish to operate the quantum computer such that $ (\delta
X)_{\mbox{\small int}}$ is much smaller than the
linear spread. To ensure this, it is enough to have an interaction
such that the system supports two
time-scales, $t$ and $\epsilon t$ where $\epsilon $ is an adiabaticity 
parameter. In such a situation,
the diffusion sets in \cite{srj1,srj2} at times of the order 
of ${1 / \epsilon}$. Thus, our arguments will
hold if the decoherence times are much lesser than $\epsilon ^{-1}$. 
In practice, this will always hold. Also, recently it has been shown that the
decoherence time (which is same as life time or running time)
is $(\Lambda - h_{KS})^{-1}$ for complex (chaotic) systems which decay, 
where $\Lambda$ and $h_{KS}$ are respectively the Lyapunov exponent 
and the Kolmogorov-Sinai entropy of the
fractal repeller containing all the periodic orbits residing in the 
system \cite{srj3}.

The minimum value of the spread occurs for $(\delta X(t))^2 \sim
(\delta X)^2 \sim
\frac{\hbar t}{ M}$. Hence, if the computation takes a time $T$ before
the quantum computer
decoheres, the resolution one can get for the spread of the wave
packet is limited by ``standard quantum
limit'', i.e., $(\delta X)_{SQL} \sim  \sqrt{ \frac{\hbar T}{ M}}$. 
Therefore, we have
\begin{equation}
\delta X  \ge  \sqrt{ \frac{\hbar T}{ M}}.
\end{equation}
The total time $T$ is also identified as the life time of a computer, 
because in order to do any useful computation in a quantum computer 
the system should not decay into some `unwanted'
states. Next, we recall `Wigner's clock' argument for a general
quantum mechanical system. Wigner \cite{epg} has argued that if a
quantum device can distinguish time-intervals to within an accuracy of $\tau$,
then the wave packet of the system will be limited to a length scale
$\delta X \le c \tau$. This
means that the wave packet of the quantum computer is limited within
the same length scale. By
combining the  above two, we have the following inequality for 
these dimensionless quantities
\begin{equation}
\frac{M c^2 T}{\hbar}  \ge \frac{T^2}{\tau^2 }.
\end{equation}

Now, we bring another important empirical observation from 
particle physics. In a series of papers a 
unifying formalism of elementary particles was
developed that remarkably relates their masses and life-times 
\cite{ram,rams,ram1,sr,rs}.
%Its validity has been argued using Cantor's 
%continuum theory \cite{ram}. 
If we have an elementary quantum system with mass
$m$ and life time $T$ in presence of certain force field, then 
it \cite{ram,rams,ram1,sr,rs} has been empirically shown that the 
dimensionless quantity constructed from $m, T, c,$ and
$\hbar$ obey the following relation
\begin{equation}
\frac{ m c^2 T }{\hbar}  = \frac{2^n}{n},
\end{equation}
where $n$ is an integer that characterises the {\em strength and type of
physical interactions present}. Multiplying both sides with ${\cal N}$
we have $\frac{ M c^2 T }{\hbar}  = {\cal N} \frac{2^n}{n}$,
where ${\cal N}$ denotes the number of units of mass $m$ placed in making
the quantum device. The total
mass $M$ of the quantum computer will be given by 
$M \simeq  m {\cal N}$. The corrections
due to binding energy is neglected here. The above relation has been 
substantiated for various elementary particles that exists in nature 
as well as for nuclei \cite{ram,rams,ram1,sr,rs}. For
example, using mass-life time relation it was shown that \cite{sr} 
the life time of proton is $\sim 5.33 \times 10^{33}$ years with 
$n= 225$. This approach is unique because it incorporates two discrete
quantities together in a way that is independent of choosing a
particular unit of mass and time.

By combining Wigner inequality (3) and the empirical, but exact relation 
(4) we obtain an inequality
\begin{equation}
\frac{T^2}{\tau^2}  \le {\cal N} \frac{2^n}{n}.
\end{equation}
We notice that $\frac{1}{\tau} = \nu$ is the clock rate of the
computer which is same as number of operations per bit per unit time
\cite{sl,ng}. Moreover, $T$ is the total running time
of a computer before it decays (in other words for practical purposes 
before the quantum computer decoheres). $T/\tau=I$ is the maximum number 
of steps that a computer can sustain in information processing.
On defining $v= I \nu$, where $v$ is
nothing but the number of logical operations that one
can perform per unit time in a quantum computer, it is actually the 
speed of a computer. Therefore, we have an upper bound
on the product of speed and running time of a quantum computer given by
\begin{equation}
v~T  \le {\cal N} \frac{2^n}{n }.
\end{equation}

Alternatively, we can interpret the above bound by saying that
fundamental forces also can set a limit on the number of steps $I$ given by
\begin{equation}
I  \le  \sqrt{ {\cal N} \frac{2^n }{n}}.
\end{equation}

The most crucial aspect of the above relation is that the rhs of above 
equation is a function of an integer
that characterises the type of interaction present in the system. This 
suggest that one can label all quantum
computer according to their speed, running time and physical
interaction and assign an index $n$. 
The above relation can be interpreted as a complimentary relation
between speed and running time of a 
quantum computer.  For a given interaction, the more the speed is, the 
less time it takes for doing a 
computation. It is also clear that  more the $n$ is, more  can we 
raise the upper bound for speed for 
a given running time. 
Quantum systems subjected to strong interaction correspond to values of $n$
less than $10$. If the interaction is of electromagnetic in origin then 
the value of $n$ lies
between $10$ and $40$. For weak interaction the values of $n$ can be
more than $40$. However, for gravitational interaction the value of
$n$ is not yet established because we still do not have a full quantum 
theory of gravity nor do we have enough data about particles decaying
through gravitational interaction. All present day proposals 
manipulate quantum states using electromagnetic interaction. If we
insert the value of $n$ lying between $10$ to $40$, our bound suggests 
that the number of coherent steps that a physical system can sustain with
electromagnetic interaction can vary between $I \sim 10 \sqrt{\cal N}$
to  $10^6 \sqrt{ \cal N}$.
Interestingly, in a universe if one makes a quantum computer with protons 
then the maximum number of steps $I$ goes as $I \sim 10^{36} \sqrt{ \cal N}$.
Therefore, proton based quantum computers could process maximum amount of
information in principle.

In standard model of quantum computation one assumes that it is
possible to do quantum error correction by adjoining additional qubits
(for encoding a single qubit in many qubits such as $5$-bit code or
$9$-bit code) \cite{nc}. This is done typically to protect quantum 
information from undergoing decoherence, thus increasing the life time 
of a computer. Since our arguments hold for any general
composite quantum system, it is expected to hold also under the
situation when one attaches additional quantum systems. In fact, one 
can see from (6) that quantum computers that uses additional qubits 
will enhance the upper bound on speed and time which is indeed
expected. Thus our limit holds when one takes into account robust 
quantum computation and quantum error correction.
Further, we will show that our limit gives an upper bound on the quality of 
coherence of a qubit. If one defines the quality factor of coherence for a
qubit as $Q= \pi \nu T$, then it has been known that to surpass active
decoherence mechanism we need $Q$ to be larger than 
$10^4 \nu t_{\rm op}$, where $t_{\rm op}=1/v$ is the time taken for an
elementary operation \cite{preskill}. From (6) one can show that the 
quality factor of a qubit obey the following inequality 
\begin{equation}
Q \le \frac{\pi 2^n}{n} \nu t_{\rm op}.
\end{equation}
For example, in the case of electromagnetic
interaction $Q$ should be less than (roghly) $10^{12}  
\nu t_{\rm op}$. Thus, our limit (8) can tell us what type of interaction
can give the desired quality factor for a qubit. It may be worth
mentioning that a recent experiment reports \cite{vion} that a qubit
in a superconducting junction circuit has been designed with a quality 
factor of quantum coherence $Q \sim 2.5 \times 10^4$.
In principle one can go still beyond as suggested here.

Before concluding, we can argue that the entropy of a quantum computer will be
limited by the type of interactions present within the system. Recall that
\cite{ng} $I$ can also be regarded as the amount of information ${\cal
I}$ that can be registered by the computer (apart from factors like $\ln
2$). Since $S = k \ln 2 {\cal I}$, where $k$ is Boltzmann constant, using (7)
one can show that the entropy of the computer will be limited by
$S  \sim  \sqrt{ {\cal N} \frac{2^n }{n} }$
Thus depending on the strength of interaction, the number of 
available states to a quantum system is also limited.

To end with, we have shown that the laws of physics suggest that 
there exist a limit on the product of speed and running time of a quantum 
computer. The ultimate limits on the performance of any quantum
computer are governed by the physical processes in the universe 
and we know that all the processes are described by these four 
fundamental interactions.  Our bound is able to grasp this 
feature succinctly. We hope that  this simple, yet novel result based  
partly on exact and partly on empirical relation, will throw new
light in designing future quantum computers built from elementary particles 
with desired speed and life time.
Further, our result suggests that we can assign an index
integer `$n$' to a class of computers based on its coherence quality that
depends on fundamental forces of nature. In addition, following Lloyd 
\cite{seth,seth1} if we view our universe as a computer then our bound will
be important in deciding the running time and speed of the computation that
universe is performing.

\vskip .5cm

\noindent
{\bf Acknowledgements:} AKP and AM thank H. D. Parab for useful discussions.

%\end{math}

\end{document}